\documentclass[twocolumn,showpacs,preprintnumbers,amsmath,amssymb]{revtex4}
\usepackage{graphicx}% Include figure files
\usepackage{dcolumn}% Align table columns on decimal point
\usepackage{bm}% bold math

\begin{document}
%\preprint{hep-ph/0210xxx}

\title{\Large ~~~~\\ Next-to-leading order QCD corrections \\ 
 to one hadron-production \\ in polarized $pp$ collisions at RHIC}
\author{\large Daniel de Florian \footnote{Partially supported by Conicet and Fundaci\'on Antorchas} }  
\email{deflo@df.uba.ar}
\affiliation{
 Departamento de F\'\i sica\\  
                       Facultad de Ciencias Exactas y Naturales \\
                       Universidad de Buenos Aires \\
                       Pabell\'on I, Ciudad Universitaria \\ 
                       (1428) Capital Federal \\
                       Argentina }
\date{\today }

\begin{abstract}
We calculate the next-to-leading order QCD corrections to the  spin-dependent cross section for single-inclusive hadron 
production in hadronic collisions.
 This process will be soon studied experimentally at RHIC, providing a
  tool to unveil the polarized gluon distribution $\Delta g$.
We observe a considerably improvement in the perturbative stability
for both unpolarized and polarized cross sections. The NLO corrections are found to be non-trivial, resulting in a reduction of the asymmetry.
\end{abstract}

\maketitle

%
%%%%%%%%%%%%%%%%%%%%%%%%%%%%%%%%%%%%%%%%%%%%%%%%%%%%%%%%%%%%%%%%
\section{Introduction}
\label{sec1}
%%%%%%%%%%%%%%%%%%%%%%%%%%%%%%%%%%%%%%%%%%%%%%%%%%%%%%%%%%%%%%%%
%
In the last years measurements of the spin asymmetries 
$A_1^N$ ($N=p,n,d$) in longitudinally polarized deep-inelastic scattering 
(DIS) have provided much new information on the spin structure of the 
nucleon \cite{data}. Recent theoretical leading order (LO) and
next-to-leading order (NLO) 
\cite{bfr,dss,ds,aac,grsv,soffer,leader,blumlein}
 analyses of the data
sets demonstrate, however, that these are not sufficient to accurately
extract the spin-dependent quark ($\Delta q = q^{\uparrow}-q^{\downarrow}$) 
and gluon ($\Delta g=g^{\uparrow}-g^{\downarrow}$) densities of the nucleon. 
This is true in particular for $\Delta g(x,Q^2)$ since it contributes to 
DIS in LO only via the $Q^2$-dependence of $g_1$ (or $A_1$) 
which could not yet be 
accurately studied experimentally. As a result of this, it turns out 
 that the $x$-shape  of $\Delta g$ seems to be hardly 
constrained at all by the DIS data, even though a tendency towards a 
fairly large positive {\em total} gluon polarization, $\int_0^1 \Delta 
g(x,Q^2=10 \; \mbox{GeV}^2) dx \sim 1$, was found \cite{bfr,dss,ds,aac,grsv,leader,blumlein}.
The measurement of $\Delta g$ thus remains one of the most interesting 
challenges for future spin physics experiments. 
In selecting suitable processes for a determination of $\Delta g$,
it is crucial to pick those that, unlike DIS, have a direct gluonic
contribution already at the lowest order. Here, one thinks in the first 
place of high-$p_{T}$ reactions in nucleon--nucleon collisions, which
have been tremendously important in the unpolarized case to constrain
the unpolarized gluon density.

At the moment, the most eagerly awaited experimental tool for the `spin 
physics' community is the RHIC\cite{rhic} collider at BNL,
 at which first runs in a 
proton--proton mode with longitudinally polarized beams are expected to be
accomplished this year. 
The center-of-mass energy
for these $pp$ collisions will be ranging between 200 and 500 GeV,
with luminosities (rising with energy) between 320 and 800 pb$^{-1}$,
respectively. One expects about $70 \%$ polarization for each beam. 
Such conditions look extremely favorable for studying the spin asymmetries
for all kinds of high-$p_{T}$ $pp$ processes that are sensitive to 
the gluon density, such as hadron, jet, prompt-photon, or heavy-flavor production.
Hadrons and jets could be {\em the} key
to $\Delta g$: at $\sqrt{S} =200(500)$~GeV, hadrons(and jets) will be
extremely copiously produced, and the corresponding cross-sections will show a strong 
sensitivity to $\Delta g$ thanks to the dominance of the 
$gg$ and $qg$ initiated subprocesses in some kinematical ranges.
While the STAR detector permits a clear determination of jets, in the case of PHENIX, where the limited coverage in rapidity does not allow to observe jets, hadrons (and especially pions) can be used as jet surrogates \cite{goto,werner}.

In order to make reliable quantitative predictions for a high-energy process,
it is crucial to determine the NLO QCD corrections
to the Born approximation. Quite in general, the key issue here is to check
the perturbative stability of the process considered, i.e. to examine
the extent to which the NLO corrections affect the cross sections and
(in spin physics) the spin asymmetries relevant for experimental measurements.
Only if the corrections are under control can a process that shows 
good sensitivity to, say, $\Delta g$ at the lowest order be regarded 
as a genuine probe of the polarized gluon distribution and be reliably 
used to extract it from future data.
Furthermore, the inclusion of extra partons in the NLO perturbative calculation also allows to improve the matching between the theoretical calculation and the realistic experimental conditions.

The calculation of the NLO QCD corrections to hadron production by polarized
hadrons is the purpose of this paper. Such a calculation needs the
one-loop $2\to 2$ and tree-level $2\to 3$ polarized  amplitudes as input. Fortunately, these
amplitudes are already known \cite{MatEl,wu}. Furthermore, several
independent methods to calculate any infrared-safe quantity in any kind of
hard unpolarized collision are at present available in the  
literature~\cite{GGK,FKS,CS}. The formalism of Ref.~\cite{FKS}
has been used in Ref.~\cite{Jets97} to construct a Monte Carlo code
that can calculate any jet infrared-safe observable
in hadron--hadron unpolarized collisions and extended,  in Ref.~\cite{jetpol}, to the polarized case (photo-production of jets has been studied in \cite{jetpol-photo}).
 In the present paper, we 
extend the method of Refs.~\cite{FKS,Jets97}  to the case of 
one single hadron inclusive observables.
 As a result, we will 
present a customized code, with which it will be possible to calculate 
any infrared-safe quantity corresponding to one-hadron production 
 to NLO accuracy, for both polarized and unpolarized collisions. It is worth noticing that the code presented here can  be used to compute the `resolved' contribution to  hadron photo-production as well. This complements the calculation of the `direct' component presented in Ref.~\cite{hadron-photo}.

This paper is organized as follows: in section \ref{sec2} we describe the formalism and the main ingredients of the calculation. In section \ref{sec3} we study the perturbative stability of the NLO results by looking at the scale dependence and $K$-factors. In section \ref{sec4} we present the results for the asymmetries at RHIC and study the possibility of extracting $\Delta g$ from $pp\rightarrow \pi$ data. 
Section \ref{sec5} contains the conclusions.

%%%%%%%%%%%%%%%%%%%%%%%%%%%%%%%%%%%%%%%%%%%%%%%%%%%%%%%%%%%%%%%%
\section{Formalism and Main Ingredients}
\label{sec2}
%%%%%%%%%%%%%%%%%%%%%%%%%%%%%%%%%%%%%%%%%%%%%%%%%%%%%%%%%%%%%%%%
  
The factorization  theorem~\cite{CSS} allows to write the cross section for the one hadron production in hadronic collisions as
\begin{eqnarray}
\hspace{-2cm} 
\label{eq:fact}
&&{d \sigma^{pp\rightarrow h X}} = \hspace*{-0.2cm}
\sum_{f_1,f_2,f} \int \hspace*{-0.1cm} 
dx_1\, dx_2\, dz \,\,{f_1^{H_1}} (x_1,\mu_{FI}^2)\,\, {f_2^{H_2}} (x_2,\mu_{FI}^2)  \nonumber \\ &&
\times {d{\hat{\sigma}}^{f_1 f_2\rightarrow fX'}} 
(x_1\, p_1,x_2\,p_2,p_{h}/z,\mu_{FI},\mu_{FF},\mu_R)\,\,
\nonumber \\ &&
\times 
 {D_f^{h}} (z,\mu_{FF}^2) \, ,
\end{eqnarray}
were  $H_1$ and $H_2$ are the colliding particles with momentum $p_1$ and $p_2$, respectively,  $h$ is the outgoing hadron with momentum $p_h$ and the sum in Eq.(\ref{eq:fact}) runs over all possible  initial and final partonic states.
The parton distributions $f_i^{H_i}$ are evaluated at the factorization scale $\mu_{FI}$, the fragmentation functions at the scale $\mu_{FF}$ and the coupling constant, appearing explicitly in the perturbative expansion of the partonic cross section, at the renormalization scale $\mu_{R}$.
The analogous of  Eq.(\ref{eq:fact}) for the polarized cross section is obtained  by replacing the parton distributions and the partonic cross section by its polarized expressions, $\Delta f_i^{H_i}$ and 
${d{\Delta \hat{\sigma}}^{f_1 f_2\rightarrow fX'}}$, respectively.

As usual, the (longitudinally polarized) asymmetry is defined by the ratio between the polarized and unpolarized cross sections
\begin{eqnarray}
{A}_{LL}^{h}=\frac{d\Delta\sigma}{d\sigma}\, .
\label{asypt}
\end{eqnarray}

The method we have used for computing the NLO corrections and implementing the results in a Monte-Carlo like code has been introduced and extensively discussed in Refs.\cite{FKS,Jets97}, and in Ref.~\cite{jetpol}. We refer the reader to those references  for details. It is worth noticing that
 with one single code we can compute unpolarized and polarized cross sections, since the structure of the corrections is the same in both cases.

As stated in the introduction, there is hardly any experimental information 
on the spin-dependent gluon density $\Delta g$ at present. In contrast, 
the quark densities are far better constrained by the existing data from 
inclusive polarized DIS. This is particularly true for the polarized $u$ and $d$
quark densities, which come out rather similar in all theoretical analyses 
performed so far. The spin-dependent sea-quark distributions seem less well
constrained; however, they are of minor importance for our 
studies.

%%====================================
\begin{figure}
\includegraphics[width=9.2cm]{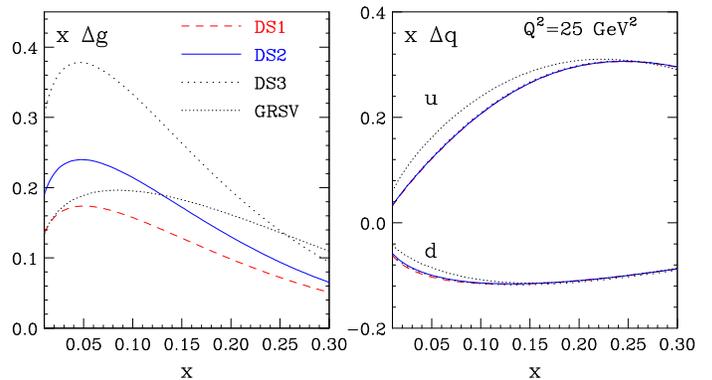}
\caption{\label{fig:pdf}{\em   The polarized gluon (left) and  quark densities (right), as
   given by the  NLO parametrizations that will be used in this paper,
   at the scale $Q^2 = 25\ \mathrm{GeV}^2$. The patterns for the quark
   densities ($\equiv \Delta q+\Delta \bar{q}$) are the same as those used for the gluon.
 In the case of DS we
 show only the results for the $+$ scenario, the ones for the $-$ scenario are identical. }}
\end{figure}
%%====================================    

In our phenomenological analysis we will try to cover as much as possible 
of the wide range of polarized parton densities allowed by the present DIS  
data: this is especially relevant for the gluon density, as we will
discuss below. For the polarized  parton densities, we will  use
the six sets of Ref.~\cite{ds} (DS1$\pm$, DS2$\pm$, DS3$\pm$), obtained by 
constraining the first moment of the polarized gluon densities in three
different ways. The first moment of the gluon of DS3 is the largest and
about three times larger than for DS1, which has the smallest 
gluon. The index $\pm$ refers to the different extremes assumptions done on the sea quark densities $\Delta \bar{u}= \pm \Delta \bar{d}$. 
Furthermore, we will also use the  `standard' set of Ref.~\cite{grsv} (from now on, 
referred to as GRSV), for comparison.
All these distributions are available at both LO and NLO, the latter 
corresponding to the $\overline{{\mbox{MS}}}$ scheme used also in our 
calculation of the NLO partonic cross section.

In this work, we will mostly concentrate on  the phenomenology of pion production at RHIC with a center-of-mass energy of $\sqrt{S}=200$ 
GeV and  transverse momenta in the region of 
$4 \ {\rm GeV} < p_{T}< 20$ GeV. This implies that the polarized
parton distributions will mainly be probed in  the $x$-range
$0.05 \lesssim x \lesssim 0.3$ and at typical scales  $Q^2$ of the order
of 25 GeV$^2$. Figure~\ref{fig:pdf} shows the NLO polarized  quark
and gluon densities of the  different sets we are going to use. It
becomes clear that there is indeed a wide range of possible gluon
distributions compatible with present polarized DIS data. As mentioned
earlier, in the quark sector, most of the distributions are very
similar. The spin-dependent sea-quark densities (not shown in
Fig.~\ref{fig:pdf}) differ more strongly among the various sets, but have
 a  small impact on the cross sections in this kinematical
region. In conclusion, since the
variations in the quark sector are much smaller than the ones for gluons, 
we can expect that any differences between predictions for the polarized 
cross sections (or asymmetries) that are found when using different 
polarized parton density sets, are to be attributed to the 
sensitivity of the observable to $\Delta g$. 

% Furthermore by using the 
% $\pm$ 'scenarios' for the antiquark densities provided by the DS sets,
% it will be possible to study the sensitivity of the observables on the quark 
% sea distribution. 

The size of radiative QCD corrections to a given unpolarized hadronic
process is often displayed in terms of a `$K$-factor' which represents the
ratio of the NLO over LO results. In the calculation of the numerator of
$K$ one obviously has to use NLO-evolved parton densities. As far as the
denominator is concerned, a natural definition requires the use of
LO-evolved parton densities. 
In the polarized case, a problem arises for such a definition. This is particularly true when the cross section is dominated by the gluon distribution: Since the data hardly constrain the gluon density, very different 
results for $\Delta g$ can emerge if the fit is performed at LO or at NLO. Therefore, the `$K$-factor' for one hadron production, defined using LO parton densities in the denominator could be artificially large or small  merely 
from the fact that the gluon is at present so ill-constrained, and not because of a `genuine' effect from higher order partonic corrections.

In order to avoid this problem, we will follow the definition introduced to perform the analysis of the results for jet production in polarized collisions at NLO \cite{jetpol}, i.e., by defining the `$K$-factor' as the ratio between the NLO and the `Born' cross section, where the latest corresponds to the use of NLO-evolved parton densities (and 2-loop expression for $\alpha_s$) when evaluating the 
 lowest-order partonic cross sections in the denominator.
Nevertheless, it is important to remember that the `$K$-factor' is {\it not} a physical quantity 
%(and it is actually very much scale dependent)
 and just provides a number to `quantify' the effect of the higher order corrections.

In the unpolarized case, we will use the GRV98 parton distributions 
\cite{grv98}. Differences of the order of $10\%$  in the cross section are observed when the more recent CTEQ \cite{cteq} or MRST\cite{mrst} distributions are used. Nevertheless, since both  DS and GRSV sets of polarized distributions used GRV98 as the reference set in the unpolarized case, either to impose 
positivity constraints or to define the measured asymmetries, we will restrict the analysis to the GRV98 set.

The other input needed in Eq.(\ref{eq:fact}) corresponds to the fragmentation functions (FF). Several sets of FF have been released in the last years \cite{bkk,kretzer,kkp,bfgw}, as the result of QCD analyses of all $e^+e^- \rightarrow h$ available data. As it is well known, $e^+e^-$ data can provide information the combination $D_q+D_{\bar{q}}$ in the quark sector and  only a weak constrain on the gluon fragmentation function, which enters at NLO or through the $Q^2$ evolution. 
A comparison of two different analysis can be found in \cite{kretzer}. In particular, large differences are observed between the BKK\cite{kkp} and Kretzer\cite{kretzer} distributions for the gluon in the whole range of $z$ and in the quark case at large $z$ (see Fig.7 of \cite{kretzer} for details).

The lack of separation between `favored' and `unfavored' quark fragmentation functions, and the large uncertainties on the distributions result on the reduction of the predictive power of pQCD calculations in  hadronic collisions. Besides the large differences than can arise in the computed cross section when different sets of FF are used, it is not possible to compute observables separately for, say, $\pi^+$ and $\pi^-$: only cross sections for the sum 
$\pi^+ + \pi^-$
 can be computed with the FF obtained from fits to $e^+e^-$ data.

An approach to solve the last problem is attempted in Kretzer's FF: 
a $(1-z)$ suppression for the `unfavored' fragmentation functions, with respect to the `favored' ones, is assumed at the low initial scale $\mu_0^2=0.4 $ GeV$^2$. Within this assumption, a full separation between quark flavors is provided, showing  a  good agreement with the available semi-inclusive DIS data.  
Furthermore, the fit by Kretzer relies on the same values of $\Lambda_{QCD}$ used by the GRV98, DS and GRSV sets, allowing to perform a fully consistent analysis in the case of hadronic collisions. 
Therefore, we will present results for hadron production at RHIC using the set of fragmentation functions presented in Ref.\cite{kretzer}. 

 Finally, it should be noticed that an analytical NLO calculation exists for the case of unpolarized hadronic collisions \cite{aversa} since long time. 
We have compared numerically to the code presented in \cite{aversa} finding excellent agreement. This provides a very important test on our calculation and, furthermore, 
 on the structure of the code that we will also use  to compute the polarized cross section.
 
%%%%%%%%%%%%%%%%%%%%%%%%%%%%%%%%%%%%%%%%%%%%%%%%%%%%%%%%%%%%%%%%%%%%%%%  
\section{Perturbative stability}  
\label{sec3}
%%%%%%%%%%%%%%%%%%%%%%%%%%%%%%%%%%%%%%%%%%%%%%%%%%%%%%%%%%%%%%%%%%%%%%%  

A reliable error estimate on our NLO results requires some knowledge on the
size of the uncalculated higher-order terms. 
 The best we can do, before higher-order terms are computed,
  is to study  the dependence of the full NLO  results
on the renormalization and factorization scales.

Although physical
observables are obviously independent of $\mu$, theoretical predictions do
have such a dependence, arising from the truncation of the perturbative
expansion at a fixed order in the coupling constant $\alpha_s$. A large
dependence on the scales, therefore, implies a large theoretical uncertainty. 
In order to show how the scale dependence is substantially reduced once
the next-to-leading order corrections are included we will compare to the Born result (with a very similar scale dependence to the LO one).
 For some observables it might occur that there is a partial cancellation in the scale dependence when all scales are varied simultaneously, either at NLO or at the Born level, and a larger dependence is observed when the scales are varied independently. Therefore, we will look at the scale dependence of the cross section when the scales are varied either simultaneously or independently, with the others fixed at a default value in the last case.

%%====================================
\begin{figure}
\hspace{-0.7cm}
\includegraphics[width=9cm]{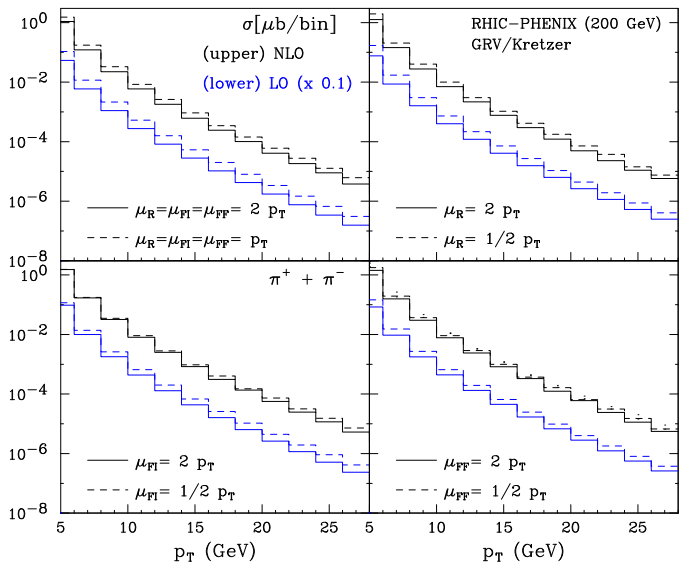}
\caption{\label{fig:scaleunpol}{\em  Scale dependence of NLO and Born unpolarized cross sections. The scales are varied around the default value $p_T$ as indicated in the text. The LO results have been rescaled by a factor $0.1$. }}
\end{figure}
%%====================================  

We begin by presenting the results for the production of $\pi^+ + \pi^-$ in unpolarized collisions at RHIC with  $\sqrt{s}=200$ GeV.
The kinematics limits of the PHENIX detector ($|\eta|<0.35$) have been imposed, but very similar results, concerning the perturbative stability, are obtained for the case of STAR (with $|\eta|<1$).

 In Figure~\ref{fig:scaleunpol} we show both NLO and Born unpolarized cross sections computed at different scales. We take as the default scales $\mu_{R}=\mu_{FI}=\mu_{FF}\equiv \mu=p_T$ and compare to the cases when the renormalization and factorizations scales are varied by a factor of 2
(up and down when the scales are varied independently, and up when  they are varied simultaneously)
 with respect to the default one. As can be observed there is a considerable reduction in the scale dependence, showing an improvement in the perturbative stability for all the scales relevant to this process.

Nevertheless, even though of the considerable improvement, it is worth
 noticing that the scale dependence is still rather large at NLO, of the order of $30\%$ or more, and certainly larger than for 
 jet production \cite{jetpol}.  Actually, the scale dependence can be much larger for other choices of parton distributions and fragmentation functions. This issue will be studied in more detail in a future publication \cite{future}.

% This can be understood  from the fact that the process under consideration 
% here is  `less inclusive' than jet production.

%%====================================
\begin{figure}
\includegraphics[width=9cm]{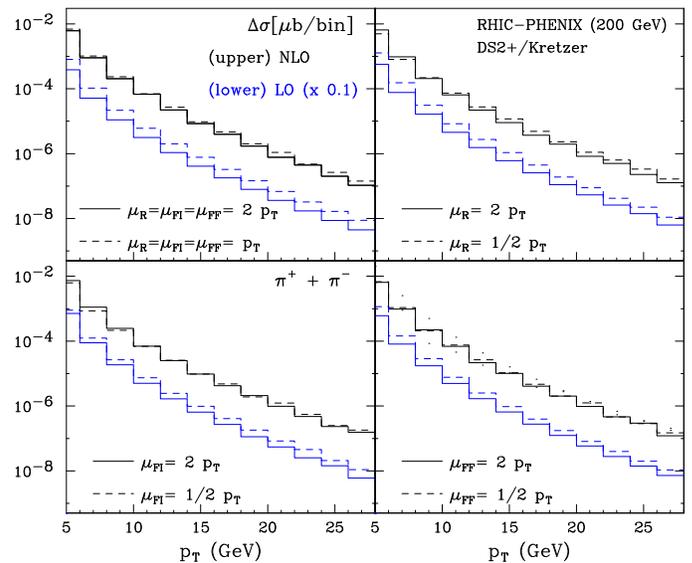}
\caption{\label{fig:scalepol}{\em  Scale dependence of NLO and Born polarized cross sections. The scales are varied around the default value $p_T$ as indicated in the text. The LO results have been rescaled by a factor $0.1$. }}
\end{figure}
%%====================================  

We move now to the polarized cross section. 
To avoid the proliferation of curves, we concentrate here on the case of the DS2+ set of polarized pdfs. 
Similar results are obtained with other distributions.

As can be observed in Figure~\ref{fig:scalepol}, the reduction is the scale dependence is even more noticeable than in the unpolarized case. A similar situation has been already observed for other observables in hadron collisions, like jets \cite{jetpol}, prompt-photon \cite{photon} and heavy quark production
\cite{heavyq}, and should  be attributed to the less singular behavior of both polarized parton densities and partonic cross sections.

\begin{figure}
\includegraphics[width=8cm]{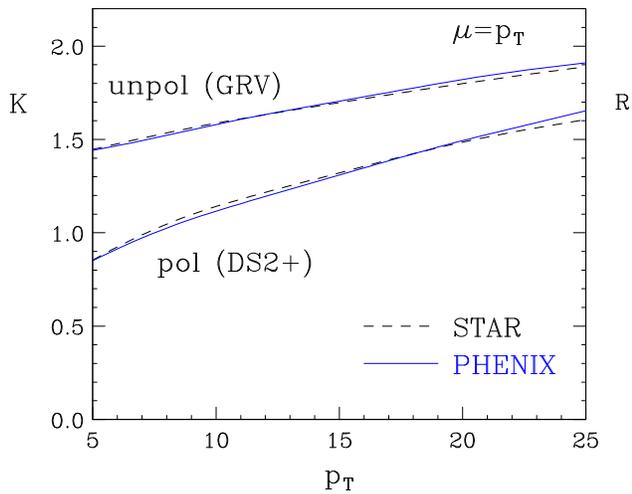}
\caption{\label{fig:kfac}{\em Unpolarized and polarized $K-$factors for the production of $\pi^+ + \pi^-$ at RHIC with $\sqrt{s}=200$ GeV.  }}
\end{figure}

We finish by presenting the $K$-factors for both polarized and unpolarized collisions in Fig.~\ref{fig:kfac}. As discussed in the previous section, $K$-factors are defined by the ratio of the NLO and Born cross sections, with all scales fixed to $\mu=p_T$. As can be observed, the corrections are larger for the unpolarized cross section than for the polarized, resulting in (about) a $30\%$ reduction of the asymmetry, for both PHENIX and STAR. The situation is very similar to what has been observed in the case of jet production in polarized hadronic collisions \cite{jetpol}.

%%====================================
%%====================================  
%%%%%%%%%%%%%%%%%%%%%%%%%%%%%%%%%%%%%%%%%%%%%%%%%%%%%%%%%%%%%%%%%%%%%%%  
\section{Asymmetries at RHIC}  
\label{sec4}
%%%%%%%%%%%%%%%%%%%%%%%%%%%%%%%%%%%%%%%%%%%%%%%%%%%%%%%%%%%%%%%%%%%%%%%  

In this section we study the phenomenology of pion production at RHIC, focusing on the possibility to extract the polarized gluon density from such a measurement.

It is actually possible to anticipate whether the process is expected to be 
sensitive to the polarized gluon distribution by looking at the relevance of the different partonic initial states in the unpolarized case, where the parton densities are
 reasonable well known.

\begin{figure}
\hspace{-1cm}
\includegraphics[width=9.5cm]{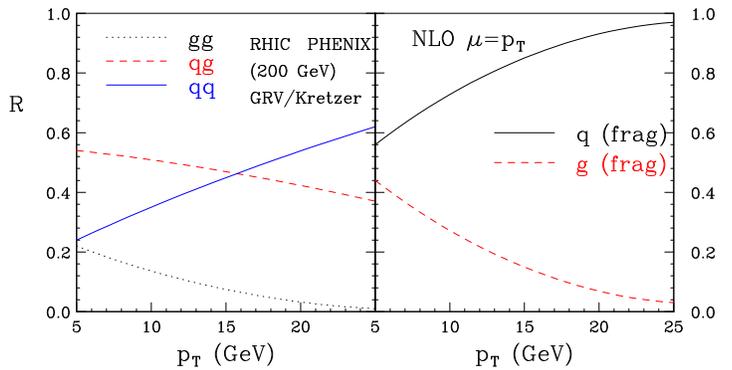}
\caption{\label{fig:ratio}{\em  Ratios of the unpolarized cross section for different combinations of initial (left) and final (right) partonic states. }}
\end{figure}

This is shown on the left-hand side  of Figure \ref{fig:ratio}, again for 
$\pi^+ + \pi^-$ production at PHENIX with $\sqrt{s}=200$ GeV. All scales have been fixed to $\mu=p_T$ and $R$ corresponds to the ratio between the contribution of the cross section by a given partonic initial state and the full one, both computed to NLO accuracy.
 As can be observed, the cross section is dominated by the $qg$ initial state at low $p_T$, with an increasing contribution of the $qq$ channel for larger transverse momentum. Considering the relevance of the gluon distribution in the unpolarized cross section, we can expect a large sensitivity on $\Delta g$ in the polarized case.
The right-hand side of Figure \ref{fig:ratio} corresponds to a similar ratio, but in this case,  to compare the relevance of the contribution of the final state (fragmenting) partons.  Except at very low $p_T$, the quark fragmentation completely dominates the cross section. It is worth noticing that the situation can change considerably when other sets of fragmentation functions are used.

%%====================================
%%==================================== 

\begin{figure}
\hspace{-1.5cm}
\includegraphics[width=9cm]{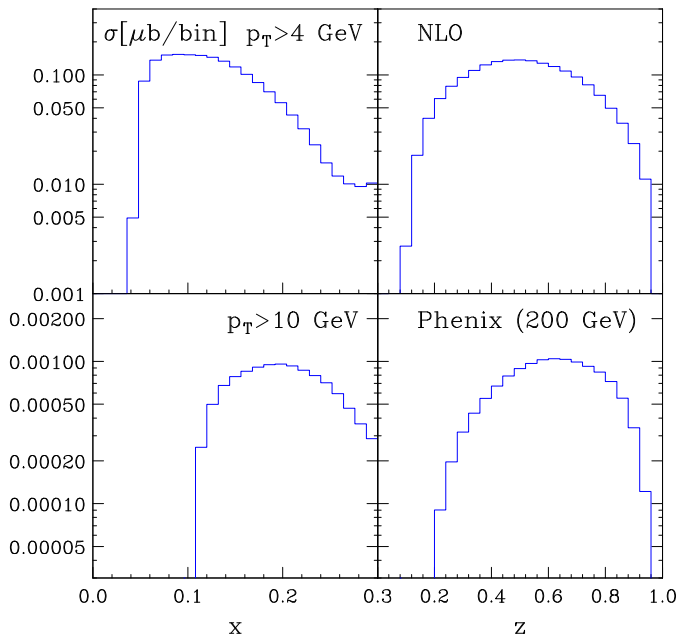}
\caption{\label{fig:xz}{\em Unpolarized cross section distributions in $x\equiv \sqrt{x_1 x_2}$ and $z$. }}
\end{figure}

Before moving to the results for the asymmetries it is interesting to find out
 the range in $x$ and $z$ covered by RHIC measurement. 
This can be 
studied by looking at the dependence of the unpolarized cross section on those 
variables, as shown in Figure \ref{fig:xz}. The results correspond to the 
NLO  integrated cross section for $p_T$ larger than 4 GeV and 10 GeV. 
As can be observed, at $\sqrt{s}=200$ GeV, the cross section is mostly
 sensitive to the parton distributions in the range $0.05<x<0.3$ and to the fragmentation functions at large $0.2<z<0.8$. It should be noticed that, unfortunately, this is the region where the uncertainties on both fragmentation and (unpolarized) parton distributions are considerably large.

We thus turn to the study of the dependence of the asymmetries on the polarized parton densities, as a way to determine the sensitivity on the polarized gluon distribution.

\begin{figure}
\hspace{-1.cm}
\includegraphics[width=9.cm]{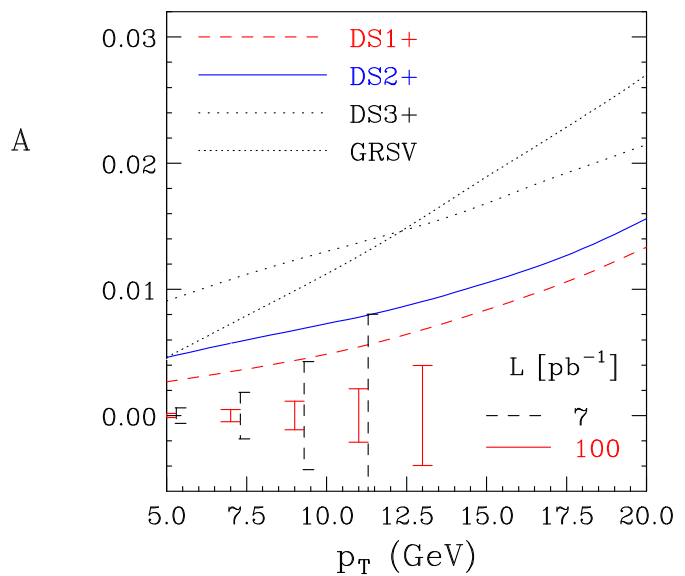}
\caption{\label{fig:asim}{\em NLO asymmetries for $\pi^+ + \pi^-$ production at PHENIX with $\sqrt{s}=200$ GeV. The `error bars' estimate the minimum observable asymmetry assuming integrated luminosities of ${\cal L}=$7 pb$^{-1}$ and 100 pb$^{-1}$. }}
\end{figure}

We begin with the asymmetry for $\pi^+ + \pi^-$ production at PHENIX with $\sqrt{s}=200$ GeV. Since the coverage in rapidity is rather small, we just concentrate on the $p_T$ distribution. The NLO results, obtained using the sets of distributions enumerated in Section 2, are shown in Fig.\ref{fig:asim}.

To estimate the minimum value of the asymmetry observable 
at RHIC, we use the well-known formula
\begin{equation}
\left({A}_{LL}^{h}\right)_{min}=
\frac{1}{P^2} \frac{1}{\sqrt{\sigma {\cal L}\epsilon}} ,
\label{Amin}
\end{equation}
where ${\cal L}$ is the integrated luminosity, $P$ is the polarization of
the beam, and the factor $\epsilon\le 1$ accounts for experimental efficiencies;
$\sigma$ is the unpolarized cross section integrated over a small
range in transverse momentum ($p_{T}$ bin). The quantity
defined in Eq.~(\ref{Amin}) is plotted as `error-bars' in Fig.~\ref{fig:asim},
for $\epsilon=1$, $P=0.7$, a $p_{T}$-bin size
of 2 GeV and two scenarios for the integrated luminosities ${\cal L}=7$~pb$^{-1}$ and $100$~pb$^{-1}$.
 From the figure we see that the asymmetry
defined in Eq.~(\ref{asypt}) is measurable if the polarized parton
densities are as described by all of the sets considered here.
Furthermore, it is clear that even with the low integrated luminosity 
expected to be collected during the first run of RHIC with longitudinally 
polarized beams ($\sim 7$~pb$^{-1}$), it will be possibly to reasonably estimate the size
of $\Delta g$ from the first $p_T$ bins. Notice that the pattern of the asymmetries computed with different polarized distributions follows closely the one showed in Fig.\ref{fig:pdf}, indicating an almost linear dependence on the gluon density and confirming the relevance of the $qg$ partonic initial state also for the polarized cross section.

%%====================================
%%====================================  

% Furthermore, the small differences between the asymmetries computed using the DS2$+$ and $-$ sets indicate that the observable is rather insensitive to the sea quark distributions. 

Smaller asymmetries are obtained for the $\sqrt{s}=500$ GeV scenario. Nevertheless, the reduction in the asymmetry is compensated by a large increase in the cross section, allowing the measurement of the asymmetry up to larger values of $p_T$.

Using charged hadrons as jets surrogates allows to investigate whether 
searching for particular hadron in the final state provides a larger asymmetry. In this paper we concentrate in the cases of $\pi^+$ and $\pi^-$, within the approximation for the fragmentation functions discussed before.
 
%%====================================
%%==================================== 

\begin{figure}
\hspace{-1cm}
\includegraphics[width=9.5cm]{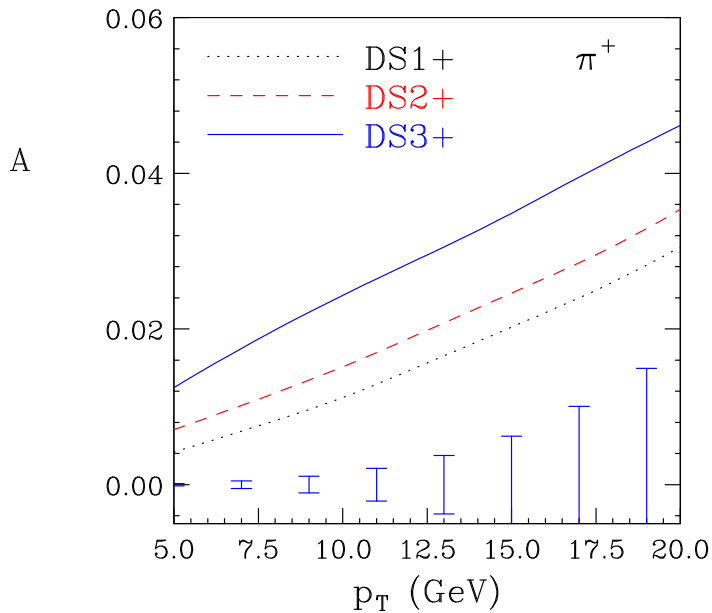}
\caption{\label{fig:asim_pimas}{\em NLO asymmetries for $\pi^+$ production at PHENIX with $\sqrt{s}=200$ GeV. }}
\end{figure}

In Fig.\ref{fig:asim_pimas} we show the NLO asymmetries computed with the DS sets for PHENIX with $\sqrt{s}=200$ GeV. The estimate for the minimum observable asymmetry has been obtained assuming an integrated luminosity of ${\cal L}=100$ pb$^{-1}$. As can be observed, the asymmetries are considerably increased (about a factor of 2) with respect to the case of  $\pi^+ + \pi^-$, while the 
 minimum observable asymmetry only increases by, roughly, a $\sqrt{2}$ factor. 
The measurement of $\pi^+$ production could provide an even more powerful tool to extract the polarized gluon distributions, after the uncertainties on flavor separation of the fragmentation functions are reduced.

\begin{figure}
\hspace{-1cm}
  \includegraphics[width=9.5cm]{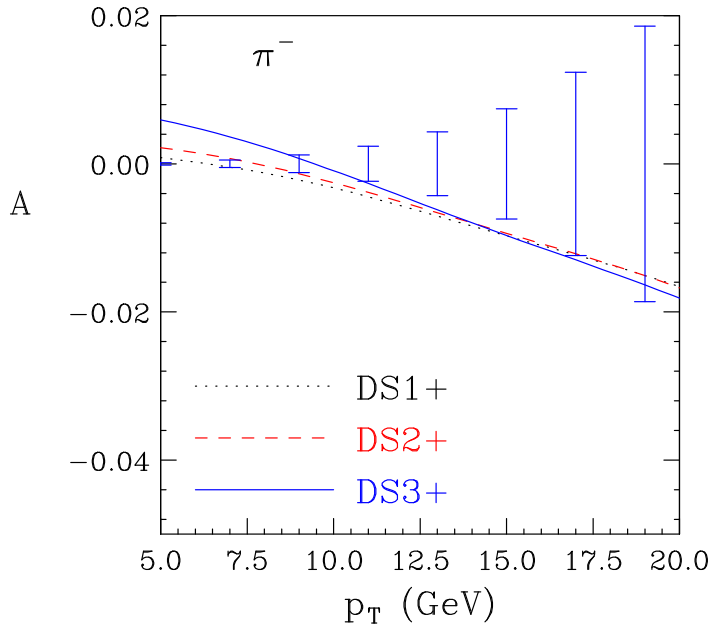}
\caption{\label{fig:asim_pimenos}{\em The same as in Fig.\ref{fig:asim_pimas} but for $\pi^-$ production. }}
\end{figure}

Finally, we move to the case of $\pi^-$ production presented in Fig.\ref{fig:asim_pimenos}. The asymmetries turn out to be much smaller than in the other cases, reaching a zero around $p_T=10$ GeV and becoming negative for larger values of the transverse momentum of the pions. Furthermore, the sensitivity of the asymmetry on $\Delta g$ is also considerably reduced, becoming very difficult to extract it from such a measurement.

It is possible to understand the differences between the asymmetries for $\pi^+$ and $\pi^-$ production on simple physical basis. 
The main contribution to the polarized cross section, at least regarding the 
terms depending  on the gluon density, appears with the following combination of parton distributions and fragmentation functions
\begin{equation}
\Delta g\, (\Delta u\, D_u^\pi + \Delta d\, D_d^\pi) \, . 
\end{equation}

In the kinematical range relevant for RHIC measurements, $\Delta u$ is found to be positive, while $\Delta d$  negative, with $|\Delta d/ \Delta u|< 1$. Furthermore, for $\pi^+$ the `favored' and `unfavored' fragmentation functions obey
$D_u^\pi > D_d^\pi$, while the opposite occurs for $\pi^-$. In the case of the  $\pi^+ +\pi^-$ production, both fragmentation functions are the same.
Therefore, for $\pi^+$ the combination $(\Delta u\, D_u^\pi + \Delta d\, D_d^\pi)$ reaches its maximum, thus increasing the dependence of the asymmetry on the polarized gluon density, while for $\pi^-$ can be zero, becoming the asymmetry dominated by other initial states, like $qq$.

Finally, it should be remarked that there are still large uncertainties on the estimates presented in this paper, mostly due to the lack of precise knowledge on the fragmentation functions. Nevertheless, those uncertainties  could be considerably reduced by using the data on hadron production collected at RHIC running with {\rm unpolarized} beams. Including these data on a {\it global} fit
 allows to define a safe procedure to extract the fragmentation functions needed to compute the measurable asymmetries in the polarized case.

%%%%%%%%%%%%%%%%%%%%%%%%%%%%%%%%%%%%%%%%%%%%%%%%%%%%%%%%%%%%%%%%
\section{Conclusions}
\label{sec5}
%%%%%%%%%%%%%%%%%%%%%%%%%%%%%%%%%%%%%%%%%%%%%%%%%%%%%%%%%%%%%%%%

%%====================================
%%==================================== 

%%====================================
%%==================================== 

We have computed  the next-to-leading order QCD corrections to the  spin-dependent cross section for single-inclusive hadron 
production in hadronic collisions.
The calculation is implemented in a Monte-Carlo like code that allows to compute any infrared-safe quantity corresponding to one-hadron production to NLO accuracy, for both polarized and unpolarized collisions.

Using the code, we have investigated the phenomenological implications for pion production at RHIC. We find that the scale dependence is considerably reduced when the NLO corrections are included.
Also the corrections are found 
to be non-trivial: $K$ factors are larger for the unpolarized cross section 
than for the polarized one, resulting in a reduction of the asymmetry at NLO.
The possibility of looking at charged pions in the final state is also 
studied, finding that $\pi^+$ and $\pi^0$ are the most convenient channels 
for the determination of the polarized gluon density at values of $x$ between 
0.05 and 0.3. 
One hadron production at RHIC provides an excellent tool
for the extraction of $\Delta g$,
 being possible to obtain a reasonable estimate of the size of the polarized gluon distribution just from the first run with longitudinally polarized beams to be performed in the next few months.
\\

%%%%%%%%%%%%%%%%%%%%%%%%%%%%%%%%%%%%%%%%%%%%%%%%%%%%%%%%%%%%%%%%
  \subsection*{Note Added}
As we completed this manuscript, we became aware of a calculation 
for the same process being  performed by B.J\"{a}ger, M.Stratmann and 
W.Vogelsang \cite{jsv}. We have compared our results numerically finding 
good agreement.
% at the percent level.

%%%%%%%%%%%%%%%%%%%%%%%%%%%%%%%%%%%%%%%%%%%%%%%%%%%%%%%%%%%%%%%%%%%%%%%  
  \subsection*{Acknowledgments}
The author is grateful to B.J\"{a}gger, M.Stratmann and W.Vogelsang for discussions and comparisons with their results, and to Mariel Gutierrez for her support.
% Este trabajo esta dedicado a la memoria de Ernesto `Che' Guevara.

%%%%%%%%%%%%%%%%%%%%%%%%%%%%%%%%%%%%%%%%%%%%%%%%%%%%%%%%%%%%%%%%%%%%%%%  
%\newpage

%%%%%%%%%%%%%%%%%%%%%%%%%%%%%%%%%%%%%%%%%%%%%%%%%%%%%%%%%%%%%%%%%%%%%%%  


\begin{thebibliography}{}  

  \section*{\Large References}

\newcommand{\np}[3]{{ Nucl. Phys.} {\bf #1}, #2 (19#3)}  
\newcommand{\pr}[3]{{ Phys. Rev.} {\bf #1}, #2 (19#3)}   
\newcommand{\zp}[3]{{ Z. Phys.} {\bf #1}, #2 (19#3)}   
\newcommand{\nc}[3]{{ Nuovo Cimento} {\bf #1}, #2 (19#3)}
\newcommand{\epj}[3]{{ Eur. Phys. J.} {\bf #1}, #2 (19#3)}
\newcommand{\jp}[3]{{ J. Phys.} {\bf #1}, #2 (19#3)}
\newcommand{\prep}[3]{{ Phys. Rept.} {\bf #1}, #2 (19#3)}

  
\bibitem{data}
 For a compilation of references to the data, see  P.J.~Mulders and
 T.~Sloan, Summary talk of Spin Physics Working Group at the $6^{\rm th}$
 International Workshop on Deep Inelastic Scattering and  QCD, Brussels
 (1998), hep-ph/9806314. 

\bibitem{bfr}
 G.~Altarelli, R.D.~Ball, S.~Forte and G.~Ridolfi,  \np{B496}{337}{97}. 
 
\bibitem{dss}
D.~de Florian, O.~A.~Sampayo and R.~Sassot,
%``Next-to-leading order analysis of inclusive and semi-inclusive  polarized data,''
Phys.\ Rev.\ D {\bf 57}, 5803 (1998).
%%CITATION = HEP-PH 9711440;%%

\bibitem{ds}
D.~de Florian and R.~Sassot,
%``Inclusive and semi-inclusive polarized DIS data revisited,''
Phys.\ Rev.\ D {\bf 62}, 094025 (2000).
%%CITATION = HEP-PH 0007068;%%

\bibitem{aac}
Y.~Goto {\it et al.}  [Asymmetry Analysis collaboration],
%``Polarized parton distribution functions in the nucleon,''
Phys.\ Rev.\ D {\bf 62}, 034017 (2000).
%%CITATION = HEP-PH 0001046;%%

\bibitem{grsv}
M.~Gluck, E.~Reya, M.~Stratmann and W.~Vogelsang,
%``Models for the polarized parton distributions of the nucleon,''
Phys.\ Rev.\ D {\bf 63}, 094005 (2001).
%%CITATION = HEP-PH 0011215;%%

\bibitem{soffer}
C.~Bourrely, J.~Soffer and F.~Buccella,
%``A statistical approach for polarized parton distributions,''
Eur.\ Phys.\ J.\ C {\bf 23}, 487 (2002).
%%CITATION = HEP-PH 0109160;%%

\bibitem{leader}
E.~Leader, A.~V.~Sidorov and D.~B.~Stamenov,
%``A new evaluation of polarized parton densities in the nucleon,''
Eur.\ Phys.\ J.\ C {\bf 23}, 479 (2002).
%%CITATION = HEP-PH 0111267;%%

\bibitem{blumlein}
J.~Blumlein and H.~Bottcher,
%``QCD analysis of polarized deep inelastic scattering data and parton  distributions,''
Nucl.\ Phys.\ B {\bf 636}, 225 (2002).
%%CITATION = HEP-PH 0203155;%%

\bibitem{rhic}
G.~Bunce, N.~Saito, J.~Soffer and W.~Vogelsang,
%``Prospects for spin physics at RHIC,''
Ann.\ Rev.\ Nucl.\ Part.\ Sci.\  {\bf 50}, 525 (2000).
%%CITATION = HEP-PH 0007218;%%

\bibitem{goto} Y. Goto, Proceedings of Spin Physics at RHIC in Year-1 and Beyond, RIKEN BNL (2001).

\bibitem{werner}W. Vogelsang, Proceedings of Spin Physics at RHIC in Year-1 and Beyond, RIKEN BNL (2001).

\bibitem{MatEl}
 Z.~Bern and D.A.~Kosower, \np{B379}{451}{92} ; \\
 Z.~Kunszt, A.~Signer and Z.~Tr\'ocs\'anyi, \np{B411}{397}{94}. 

\bibitem{wu}
 R.\ Gastmans and T.T.\ Wu, {\it The Ubiquitous Photon}, 1990, 
 Clarendon Press, Oxford.

\bibitem{GGK}
 W.T.~Giele and E.W.N.~Glover, \pr{D46}{1980}{92} ;\\
 W.T.~Giele, E.W.N.~Glover and D.A.~Kosower, \np{B403}{633}{93}.

\bibitem{FKS}
 S.~Frixione, Z.~Kunszt and A.~Signer, \np{B467}{399}{96}.

\bibitem{CS}
 S.~Catani and M.~Seymour, \np{B485}{291}{97} 
  (Erratum \np{B510}{503}{98} ).

\bibitem{Jets97}
 S.~Frixione, \np{B507}{295}{97}.

\bibitem{jetpol}
D.~de Florian, S.~Frixione, A.~Signer and W.~Vogelsang,
%``Next-to-leading order jet cross sections in polarized hadronic  collisions,'' 
Nucl.\ Phys.\ B {\bf 539}, 455 (1999).
%%CITATION = HEP-PH 9808262;%%

  \bibitem{jetpol-photo}
 D.~de Florian and S.~Frixione,
%``Jet cross sections in polarized photon hadron collisions,''
 Phys.\ Lett.\ B {\bf 457}, 236 (1999).
%%CITATION = HEP-PH 9904320;%%

\bibitem{hadron-photo}
D.~de Florian and W.~Vogelsang,
%``Next-to-leading order {QCD} corrections to inclusive hadron  photoproduction in polarized lepton proton collisions,''
Phys.\ Rev.\ D {\bf 57}, 4376 (1998).
%%CITATION = HEP-PH 9712273;%%

\bibitem{CSS}
 J.C.~Collins, D.E.~Soper and G.~Sterman, in {\it Perturbative
 Quantum Chromodynamics}, ed. A.~Mueller, 1989, World Scientific,
 Singapore, and references therein.

\bibitem{grv98}
M.~Gluck, E.~Reya and A.~Vogt,
%``Dynamical parton distributions revisited,''
Eur.\ Phys.\ J.\ C {\bf 5}, 461 (1998).
%%CITATION = HEP-PH 9806404;%%

\bibitem{cteq}
J.~Pumplin, D.~R.~Stump, J.~Huston, H.~L.~Lai, P.~Nadolsky and W.~K.~Tung,
%``New generation of parton distributions with uncertainties from global  QCD analysis,''
JHEP {\bf 0207}, 012 (2002).
%%CITATION = HEP-PH 0201195;%%

\bibitem{mrst}
A.~D.~Martin, R.~G.~Roberts, W.~J.~Stirling and R.~S.~Thorne,
%``NNLO global parton analysis,''
Phys.\ Lett.\ B {\bf 531}, 216 (2002).
%%CITATION = HEP-PH 0201127;%%

\bibitem{bkk}
J.~Binnewies, B.~A.~Kniehl and G.~Kramer,
%``Pion and kaon production in e+ e- and e p collisions at next-to-leading order,''
Phys.\ Rev.\ D {\bf 52}, 4947 (1995).
%%CITATION = HEP-PH 9503464;%%


\bibitem{kretzer}
S.~Kretzer,
%``Fragmentation functions from flavour-inclusive and flavour-tagged e+ e-  annihilations,''
Phys.\ Rev.\ D {\bf 62}, 054001 (2000).
%%CITATION = HEP-PH 0003177;%%

\bibitem{kkp}
B.~A.~Kniehl, G.~Kramer and B.~Potter,
%``Testing the universality of fragmentation functions,''
Nucl.\ Phys.\ B {\bf 597}, 337 (2001).
%%CITATION = HEP-PH 0011155;%%

\bibitem{bfgw}
L.~Bourhis, M.~Fontannaz, J.~P.~Guillet and M.~Werlen,
%``Next-to-leading order determination of fragmentation functions,''
Eur.\ Phys.\ J.\ C {\bf 19}, 89 (2001).
%%CITATION = HEP-PH 0009101;%%

\bibitem{aversa} F.Aversa, P.Chiappeta, M.Greco and J.-Ph.Guillet,
Nucl.\ Phys.\ B {\bf 327} 105 (1989).

\bibitem{photon}
L.~E.~Gordon and W.~Vogelsang,
%``Polarized and unpolarized prompt photon production beyond the leading order,''
Phys.\ Rev.\ D {\bf 48}, 3136 (1993); \\
%%CITATION = PHRVA,D48,3136;%%
A.~P.~Contogouris, B.~Kamal, Z.~Merebashvili and F.~V.~Tkachov,
%``Complete Next-To-Leading Order Corrections For Direct Photon Production By Polarized Beam And Target,''
Phys.\ Lett.\ B {\bf 304}, 329 (1993); \\
%%CITATION = PHLTA,B304,329;%%
L.~E.~Gordon and W.~Vogelsang,
%``Prompt photon production with polarized proton beams in next-to-leading order,''
Phys.\ Rev.\ D {\bf 49}, 170 (1994);\\
%%CITATION = PHRVA,D49,170;%%
A.~P.~Contogouris, B.~Kamal, Z.~Merebashvili and F.~V.~Tkachov,
%``Direct Photon Production By Polarized Beam And Target: Complete Next-To-Leading Order Corrections,''
Phys.\ Rev.\ D {\bf 48}, 4092 (1993)
[Erratum-ibid.\ D {\bf 54}, 7081 (1996)];\\
%%CITATION = PHRVA,D48,4092;%%
S.~Frixione and W.~Vogelsang,
%``Isolated photon production in polarized p p collisions,''
Nucl.\ Phys.\ B {\bf 568}, 60 (2000).
%%CITATION = HEP-PH 9908387;%%


\bibitem{heavyq}
I.~Bojak and M.~Stratmann,
%``Next-to-leading order QCD corrections to the polarized hadroproduction  of heavy flavors,''
arXiv:hep-ph/0112276.
%%CITATION = HEP-PH 0112276;%%

\bibitem{jsv} B. J\"{a}ger, M. Stratmann and W. Vogelsang, in preparation.
%arXiv:hep-ph/0209xxx.

\bibitem{future}D. de Florian, B. J\"{a}ger, M. Stratmann and W. Vogelsang, 
in preparation.

  
\end{thebibliography}
\end{document}